\documentclass[showpacs]{revtex4}
\usepackage{amsfonts}
\usepackage{amsmath}
\usepackage{amssymb}
\usepackage{graphicx}

\begin{document}

\title{Inhomogeneous light photovoltaic effect in neighboring quantum dots}
\author{Wenxi Lai}\email{wxlai@pku.edu.cn}
\affiliation{School of Applied Science, Beijing Information Science and Technology University, Beijing 100192, China}

\begin{abstract}
Photovoltaic effect of double quantum dots under nonuniform light field intensity has been studied theoretically. Comparing with the traditional p-n type photovoltaic effect, the inhomogeneous light field provides asymmetric potential creating polarization of electron number distribution in the neighboring quantum dots and furthermore gives rise to net current. Current density and efficiency of such kind solar cells are estimated to be comparable to the traditional p-n type material based solar cells. Motion of electron is described using quantum master equation around room temperature. The inhomogeneous light photovoltaic effect has potential applications for the gain of more economical solar cells.
\end{abstract}

%73.63.Kv Quantum dots
%84.60.Jt Photoelectric conversion: solar cells and arrays
%72.40.+w Photoconduction and photovoltaic effects
%73.23.Hk Coulomb blockade; single-electron tunneling

\maketitle

\begin{center}
\textbf{1. Introduction}
\end{center}

Photovoltaic effect is one of the important ways to obtain green energy, which highlights the researches and development of solar cells. Solar cell technologies are classified into three generations~\cite{Ranabhat}. First generation solar cells are based on crystalline wafer of silicon~\cite{Choubey,Bagher,Srinivas,Saga}. They have advantage of higher conversion efficiency compared to the other types of cells, but have disadvantage of high costs at the same time~\cite{Green}. Second generation solar cells are made from layers only a few micrometers thick films which is much thicker than crystalline silicon based cells~\cite{Chopra,Badawy,Razykov}. Therefore, they are also called thin film solar cells. The advantages of thin film cells are integrable, flexible and economical, however, they have drawback like the feature of low efficiency or toxicity~\cite{Ranabhat}. Third generation solar cells do not rely on traditional p-n junctions, they are made from new developed sensitizing materials, such as dye molecules~\cite{Shalom}, quantum dots (QDs)~\cite{Hu,Tutu,Sogabe} and organic polymers~\cite{Baek}. These nano materials could help us to harvest light energy at low cost and low heat emission compared with the two generation solar cells.

Actually, the third generation solar cells inherit some significant features of the first and second generation cells. For example, p-n type bilayer structure, integrability, thin film designs still play important role in recent research and development of nano material based solar cells~\cite{Hu,Tutu,Sogabe,Baek,Cheriton,Shi}. p-type and n-type materials are significant since the p-n junctions could provide potential difference to the cell system for the separation and collection of charge careers, such as electrons (negative), holes (positive)~\cite{Nozik}. Task of light is just excite electrons into higher levels of nano crystals or molecules, then electrons and holes would be separated naturally due to the asymmetric potential of p-type and n-type bilayer.

\begin{figure}
% Requires \usepackage{graphicx}
\includegraphics[width=8cm]{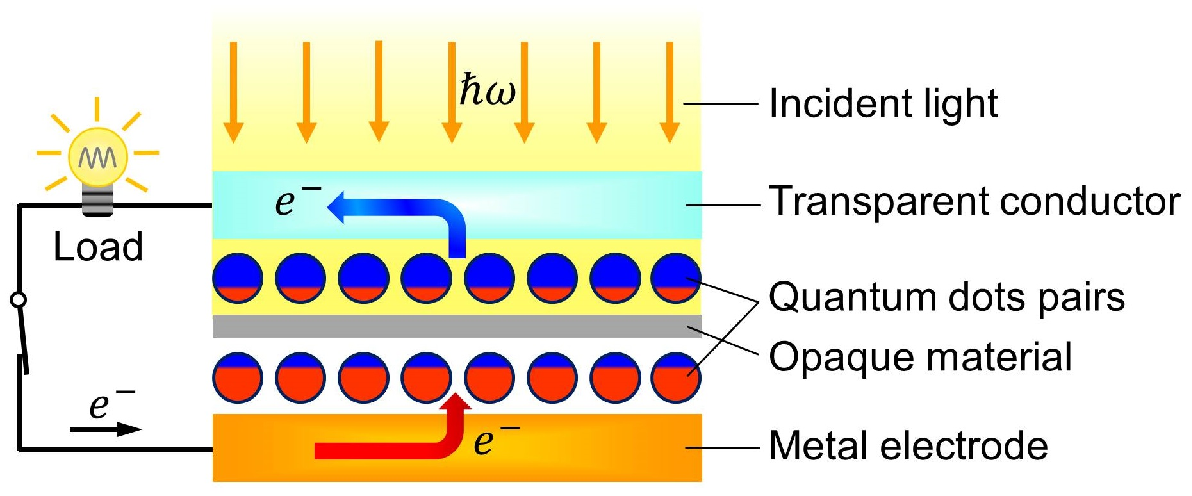}\\
\caption{(Color on line) Conceptual illustration of the inhomogeneous field QD solar cell.}
\label{figure1}
\end{figure}

In this paper, we propose a new type of photovoltaic effect based on inhomogeneous light intensity in neighboring double QDs system. With this model, asymmetric potential in p-n type bilayer of traditional solar cells could be replaced by the asymmetric field intensity of inhomogeneous light. For a definite air mass (AM), power of light transferred to the earth surface is uniform. How to obtain spatially nonuniform light field on a tiny sized area is the challenging technique of this kind photovoltaic effect. In Fig.~\ref{figure1}, we give a feasible scheme for the design of the inhomogeneous field solar cell. The configuration is symmetrically structured as conductor-QDs-conductor. A non-transparent material is sandwiched between the QDs film. The non-transparent material constructs gradient of light intensity in the QDs film.

Advantage of such solar cell is very simply structured. It can be made more economically since the p-type and n-type semiconductor materials are not required here. Furthermore, particular toxic material is not necessary for the cell. Costs of this new technology include manufacturing ultra thin QDs film and creation of light intensity difference on the size of a few QDs.

\begin{center}
\textbf{2. Open system model of the neighboring double QDs}
\end{center}

\begin{figure}
% Requires \usepackage{graphicx}
\includegraphics[width=12cm]{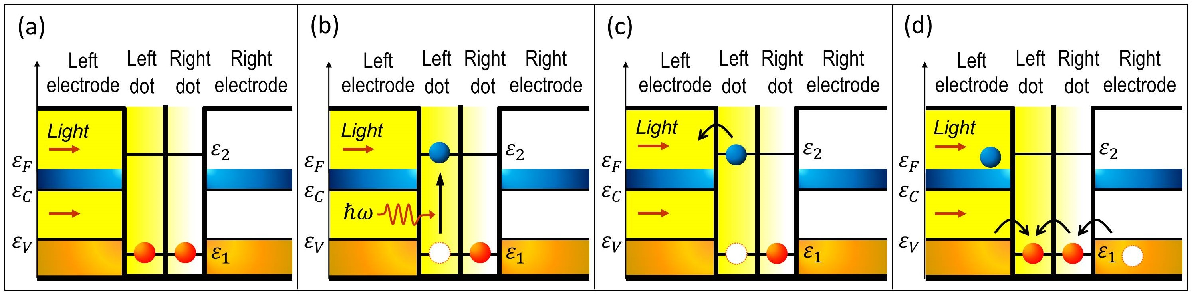}\\
\caption{(Color on line) Schematic display of inhomogeneous field double-QD photovoltaic system. (a) Initially, electrons from the valence band enter the double-QD. (b) An electron in the left QD is driven by input light and excited to its higher energy state. (c) Due to Coulomb blockade effect, the excited electron have small probability to tunnel into the right QD, as a result, it would most likely transfer into conduction band of the left electrode. (d) Electrons from the valence band tunnel into the double-QD and then the system have returned to its original state of (a).}
\label{figure2}
\end{figure}

In the following, we will describe the inhomogeneous light photovoltaic effect in neighboring double QDs system as illustrated in Fig.~\ref{figure2} (a)-(d). The double QDs system is assumed to be coupled to left (L) and right (R) electrodes with the same Fermi level $\varepsilon_{F}$. As both of the two electrodes are conductors, $\varepsilon_{F}>\varepsilon_{C}$, $\varepsilon_{V}$, where $\varepsilon_{C}$ is the bottom level of conductance band and $\varepsilon_{V}$ is the top level of valence band. The ground and excited levels of the QDs are denoted by $\varepsilon_{1}$ and  $\varepsilon_{2}$, respectively. They are required to satisfy $\varepsilon_{1}<\varepsilon_{F}<\varepsilon_{2}$ here as shown in Fig.~\ref{figure2}. Intensities of sun light in the left QD and right QD are assumed to be tunable. For convenience of numerical treatment, single-electron occupation (works in strong Coulomb blockade regime) and two levels of each QD is considered. Motion of hole is not described here since electron and hole map one by one. The total Hamiltonian $H$ can be separated into two parts for QDs $H_{dot}$ and electrodes $H_{pole}$, namely, $H=H_{dot}+H_{pole}$. Then, neighboring two QDs that coupled to external light could be described in the single-particle Hamiltonian,
\begin{eqnarray}
H_{dot}&=&\sum_{\alpha,s}\varepsilon_{s}n_{\alpha s}+\hbar\Lambda\sum_{s}(c_{L s}^{\dag}c_{R s}+H.c.)+\hbar\sum_{\alpha}\frac{\Omega_{\alpha}}{2}(c_{\alpha 1}^{\dag}c_{\alpha 2}e^{i(\omega t+\phi_{\alpha})}+H.c.).
\label{eq:dot}
\end{eqnarray}
Here, $n_{\alpha s}=c_{\alpha s}^{\dag}c_{\alpha s}$ represents occupation number operator of electron in state $s$ in QD $\alpha$ ($\alpha=L$, $R$), where, $s=1$ and $s=2$ denote the ground and excited states of an individual electron, respectively. Electrons are allowed to tunnel from a left dot to the right dot with the rate $\Lambda$, and vice versa. Each QD is coupled to the input light with the Rabi frequency $\Omega_{\alpha}$. The main frequency $\omega$ of sun light is considered in the Hamiltonian and initial phase of the light is included in $\phi_{\alpha}$~\cite{Scully}.

As conductive materials, the electrodes could be described by the energy of free electronic gas,
\begin{eqnarray}
H_{pole}=\sum_{\alpha,k}\varepsilon_{\alpha k}n_{\alpha k}+\hbar g\sum_{\alpha,k,s}[a_{\alpha k}^{\dag}c_{\alpha s}+H.c.].
\label{eq:electrode}
\end{eqnarray}
The first part of \eqref{eq:electrode} describes energy of free electrons the left ($\alpha=L$) and right ($\alpha=R$) electrodes. Operator $n_{\alpha k}=a_{\alpha k}^{\dag}a_{\alpha k}$ represents occupation number of electrons with energy $\varepsilon_{\alpha k}$ in the conductance or valance band. The third part represents coupling between electrodes and QDs. The tunneling amplitude $g$ is considered to be insensitive to the electronic states $s$ and the same for the left and right electrodes.

\begin{center}
\textbf{3. Quantum master equation and current}
\end{center}

Using the total Hamiltonian $H$ of the opened QDs system, equation of motion of electron due to the inhomogeneous light excitation can be derived under in the weak coupling limit of nano device~\cite{Lai,Scully,Delerue},
\begin{eqnarray}
\frac{\partial\rho}{\partial t}=-\frac{i}{\hbar}[V_{dot},\rho]+\mathcal{L}_{L}\rho+\mathcal{L}_{R}\rho,
\label{eq:equation}
\end{eqnarray}
which is a Lindblad form quantum master equation with the effective Hamiltonian $V_{dot}=\Delta \sum_{\alpha}n_{\alpha 2}+\hbar\Lambda\sum_{s}(c_{Ls}^{\dag}c_{Rs}+H.c.)+\hbar\sum_{\alpha}\frac{\Omega_{\alpha}}{2}(c_{\alpha 1}^{\dag}c_{\alpha 2}e^{i\phi}+H.c.)$ of QDs. The second and third term indicate the exchange of electrons between QDs and electrodes. The Lindblad super operators acting on the density matrix $\rho$ can be written as $\mathcal{L}_{L}\rho=\frac{\Gamma}{2}\sum_{s}[f(\varepsilon_{s})(c_{L s}^{\dag}\rho c_{L s}-c_{L s}c_{L s}^{\dag}\rho)+(1-f(\varepsilon_{s}))(c_{L s}\rho c_{L s}^{\dag}-c_{L s}^{\dag}c_{L s}\rho)+H.c.]$ and $\mathcal{L}_{R}\rho=\frac{\Gamma}{2}\sum_{s}[f(\varepsilon_{s})(c_{R s}^{\dag}\rho c_{R s}-c_{R s}c_{R s}^{\dag}\rho)+(1-f(\varepsilon_{s}))(c_{R s}\rho c_{R s}^{\dag}-c_{R s}^{\dag}c_{R s}\rho)+H.c.]$, respectively. Single electron tunneling rate between a dot and an electrode is given by the energy independent coefficient $\Gamma=2\pi|g|^{2}D(\varepsilon_{s})$ under adiabatic approximation ($D(\varepsilon_{s})$ is density of states of electrons in any electrode at energy level $\varepsilon_{s}$). Mean occupation number of the single-electron state with level $\varepsilon_{s}$ in any electrode is given by the Fermi-Dirac distribution function $f(\varepsilon_{s})=\frac{1}{e^{(\varepsilon_{s}-\varepsilon_{F})/k_{B}T}+1}$ at equilibrium temperature $T$ and Fermi level $\varepsilon_{F}$.

Considering single-electron transit in a QD, three states would be involved in each dot, empty state $|0\rangle_{\alpha}$, occupation of a ground state electron $|1_{1}\rangle_{\alpha}$, and occupation of an excited state electron $|1_{2}\rangle_{\alpha}$. Mean occupation number of the single-electron state $|s\rangle_{\alpha}$ in QD $\alpha$ (left or right) can be defined as $P_{\alpha 0}=tr\{c_{\alpha s}c_{\alpha s}^{\dag}\rho\}$, $P_{\alpha 1}=tr\{c_{\alpha 1}^{\dag}c_{\alpha 1}\rho\}$, $P_{\alpha2}=tr\{c_{\alpha 2}^{\dag}c_{\alpha 2}\rho\}$, respectively. They are determined by the equation of motion \eqref{eq:equation}. tr here represents trace over the $9$ double-QD states $\{|s\rangle_{L}|r\rangle_{R}$ $|$ $s$, $r$=$0$, $1_{1}$, $1_{2}$ $\}$.

Defining the total occupation number of electrons in the double-QD at time $t$ by $N(t)=\Sigma_{\alpha,s}tr\{n_{\alpha s}\rho(t)\}$, current in the unit component could be calculated using the continuity equation\cite{Davies,Jauho,Twamley}:
\begin{eqnarray}
-e\frac{d N(t)}{dt}=j_{L}-j_{R},
\label{eq:continuity}
\end{eqnarray}
where $j_{L}$ and $j_{R}$ denotes left and right sides currents of the double-QD system. According to Kirchhoff's Laws, left and right currents should satisfy $j_{L}-j_{R}=0$ in stationary circuit. Therefore, current through system can be written as $j=(j_{L}+j_{R})/2$. Here, direction of the positive current is defined to be from left to right. Substituting Eq. \eqref{eq:equation} into the continuity equation\eqref{eq:continuity}, expressions of the current $j$ can be reached in the form~\cite{Delerue}
\begin{eqnarray}
j&=&-\frac{e\Gamma}{2} \sum_{s}[f(\varepsilon_{s})(P_{L0}-P_{R0})-(1-f(\varepsilon_{s}))(P_{Ls}-P_{Rs})].
\label{eq:current}
\end{eqnarray}
Eq. \eqref{eq:current} indicates that net current $j$ in the cell is directly proportional to the polarization (the difference of mean occupation number) of the electron number distribution in the neighboring QDs. To understand the current formula \eqref{eq:current}, we may write it in the conceptual relation as $j=Q/\tau$~\cite{Puers}, where $Q=-e\sum_{s}[f(\varepsilon_{s})(P_{L0}-P_{R0})-(1-f(\varepsilon_{s}))(P_{Ls}-P_{Rs})]$ is the inversion charge difference in the double-QD system and $\tau=2/\Gamma$ is the characteristic time of an electron transit from one electrode to the other electrode through QDs.

\begin{figure}
% Requires \usepackage{graphicx}
\includegraphics[width=11cm]{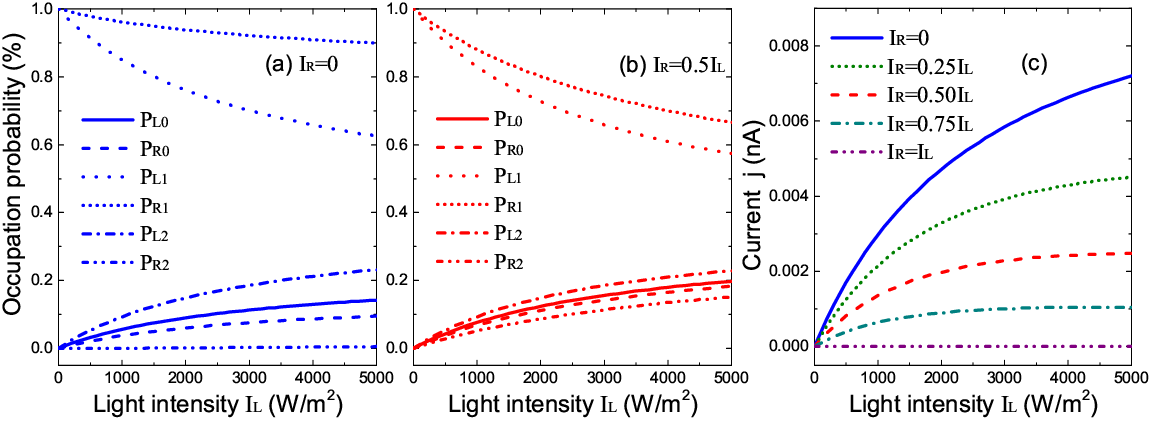}\\
\caption{(Color on line) (a) Mean occupation numbers of electrons in different states with $I_{R}=0$. (b) Mean occupation numbers of electrons in different states with $I_{R}=0.5I_{L}$. (c) Current of the double-QD unit component under given inhomogeneous fields.}
\label{figure3}
\end{figure}

\begin{center}
\textbf{4. Photocurrent of the double QDs under inhomogeneous light field}
\end{center}

Current \eqref{eq:current} can be calculated associated with the solutions of Eq. \eqref{eq:equation} under the normalization constraint $tr\rho=1$. The Rabi frequency $\Omega_{\alpha}=\frac{|\wp_{12}|E_{\alpha}}{\hbar}$ is proportional to electric field amplitude $E$ and the absolute value of dipole moment $\wp_{12}$. For the semiconductor QD, dipole moment is around $\wp_{12}=30\sim90$ D~\cite{Kamada,Stievater,Ares} for different material and dimensions. In this way, the Rabi frequency can be connected to the light intensity as $\Omega_{\alpha}=\frac{|\wp_{12}|}{\hbar}\sqrt{\frac{2c\mu_{0}I_{\alpha}}{n}}$ for $\alpha=L$ and $\alpha=R$. According to electromagnetic field theory, light intensity $I_{\alpha}$ in dot $\alpha$ is proportional to square of electric field amplitude $E_{\alpha}$, namely $I_{\alpha}=\frac{n}{2 c \mu_{0}}|E_{\alpha}|^{2}$, where $n$ is refractive index, $c$ is speed of light, and $\mu_{0}$ is permeability of vacuum. After sunlight travels through the earth's atmosphere to the earth's surface, it's intensity is reduced to nearly $1000$ W/m$^{2}$ which is called $1$ sun. Basic parameters used in this work are $\mu=70$ D, $\Gamma=\Lambda=0.5$ GHz~\cite{Sargent,Sanehira,Kamat}, $\Lambda=\Gamma$, $\phi=0$, $n=3.0$~\cite{Ravindra,Tripathy}, $T=300$ K, $\varepsilon_{e}-\varepsilon_{g}=1.77$ eV~\cite{Ranabhat,Jasim} unless some of them are taken as variables in a figure. In resonant absorption ($\Delta=0$), $\hbar\omega=1.77$ eV which is corresponding to a photon with wavelength $\lambda=700$ nm.

In Fig.~\ref{figure3}, mean occupation numbers of single-electron states in left and right QD are shown as a function of input light intensity $I_{L}$. Without incident light, $I_{L}=0$, both the two QDs are fully occupied by the ground state electrons, with $P_{L1}=P_{R1}=1$. When light is entering with different intensities in the left and right QDs, the cell begins to work. In Fig.~\ref{figure3} (a), light intensity in the left QD is increasing, at the same time, light intensity in the right QD is set to be zero. As a result, the occupation probabilities of excited electron in the left QD is much larger than that in the right QD, namely $P_{L2}\gg P_{R2}$. In addition, probability of empty occupation in the left QD is larger than that in the right QD, $P_{L0}>P_{R0}$. Corresponding current of this situation is plotted in Fig.~\ref{figure3} (c) with the same colored solid (blue) line. When light intensity in the right QD is half of the intensity in the left QD, $I_{R}=I_{L}/2$, the mean occupation number of excited electron in the right QD is remarkably increased. It decreases photocurrent as shown in Fig.~\ref{figure3} (c) with the same colored (red) line. Fig.~\ref{figure3} indicates intensity difference of light in neighboring QDs create polarization of electron number distribution and leads to net current in the opened QDs system.

\begin{center}
\textbf{5. Estimate calculations for inhomogeneous light solar cells}
\end{center}

To compare with the optoelectronic p-n junction in traditional solar cells, in Fig.~\ref{figure1} we have designed a solar cell based on the inhomogeneous light photovoltaic effect to estimate its current density and efficiency in the following. The current density $J$ of the solar cell should be calculated as
\begin{eqnarray}
J=\sigma j,
\label{eq:current-density}
\end{eqnarray}
where $j$ represents current through the single unit component and $\sigma$ denotes surface density of the unit component. Here, $\sigma$ would be taken as the sheet density of QDs. The sheet QD density commonly ranged from $10^{9}$ cm$^{-2}$ to $10^{11}$ cm$^{-2}$ which can be controlled by QD growth techniques~\cite{Li,Wigger,Kasprzak}. Here, a typical value of $\sigma=8\times10^{9}$ cm$^{-2}$ has been taken in the numerical treatments.

\begin{figure}
% Requires \usepackage{graphicx}
\includegraphics[width=11cm]{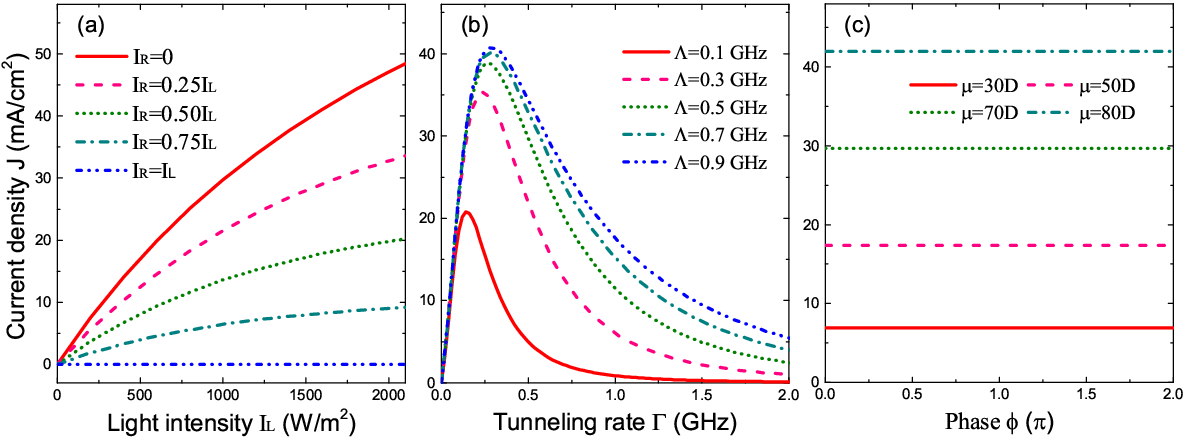}\\
\caption{(Color on line) (a) Current densities for the change of input light intensity ratio $I_{R}/I_{L}$. (b) Current densities versus electron tunneling rate $\Gamma$ under different inter-dot tunneling rates $\Lambda$. (c) Current densities as a function of light phase $\phi$ at different diploe moments $\mu$. In (b) and (c), $I_{L}=1000$ W/m$^{2}$ and $I_{R}=0$.}
\label{figure4}
\end{figure}

Considering the whole cell, current densities for different parameters have been plotted in Fig.~\ref{figure4}. Fig.~\ref{figure4} (a) reveals increase of the light intensity gradient would enhance current density for a given input light. The rates of electron tunneling $\Gamma$ and $\Lambda$ play different roles in the system as shown in Fig.~\ref{figure4} (b). When $\Gamma$ is very small, electrons are hard to exchange between electrodes and QDs, which would depress current. When $\Gamma$ is too large, polarization of electron number distribution in the neighboring QDs hard to be constructed. As a result, currents have top values for the change of $\Gamma$. In contrast, the inter-QD coupling strength $\Lambda$ always improves current. Although our results are achieved with the model using coherent light, the inhomogeneous field solar cell should work under sun light which has much short coherent time. It is demonstrated in Fig.~\ref{figure4} (c) that current in the system is insensitive to the phase $\phi$ of light. Fig.~\ref{figure4} (c) also reflects the dipole moment $\mu$ improves photocurrent due to its connection with the Rabi frequency. As manifested in the figures, current densities in the inhomogeneous field cell can be comparable to the traditional p-n type material based solar cells~\cite{Shalom,Hu,Tutu,Sogabe,Baek,Cheriton,Shi,Lee,Cao,Nozik16}.

\begin{figure}
% Requires \usepackage{graphicx}
\includegraphics[width=11cm]{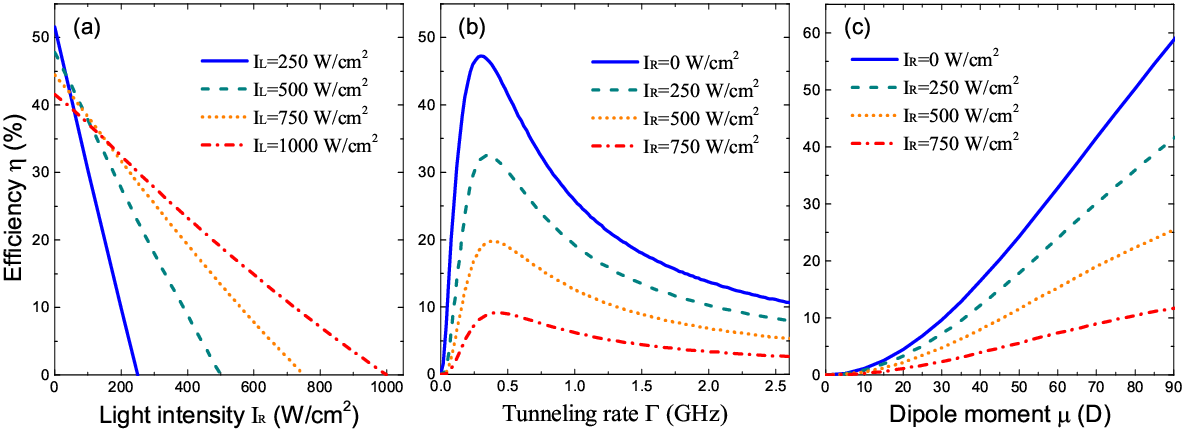}\\
\caption{(Color on line) (a) Light difference dependent cell efficiencies. (b) Cell efficiencies versus electron tunneling rate. (c) Dipole moment dependent cell efficiencies. In (b) and (c), $I_{L}=1000$ W/m$^{2}$.}
\label{figure5}
\end{figure}

Next, let us try to calculate the photocurrent efficiency. Current of electron through an area $S$ of the solar cell could be calculated as $I_{e}=JS$. Corresponding power is equal to $P_{e}=I_{e}U_{e}$, where the voltage $U_{e}$ should be related to the energy change of electron in the cell, namely the energy level difference $\varepsilon_{2}-\varepsilon_{1}$ of a QD. Therefore, we have $U_{e}=(\varepsilon_{2}-\varepsilon_{1})/e$. Power of light $P_{L}$ on the area $S$ is equal to $P_{L}=I_{L}S$. From the electronic current power and light power, we may calculate the efficiency of the inhomogeneous field solar cell through the formula,
\begin{eqnarray}
\eta=\frac{P_{e}}{P_{L}}\times 100\%.
\label{eq:efficiency}
\end{eqnarray}

Fig.~\ref{figure5} (a) shows linear relationship between light intensity difference $I_{L}-I_{R}$ in neighboring QDs and the photocurrent efficiency $\eta$. Similar to the behavior of current density, efficiency has a highest value for the electron transit rate $\Gamma$ (see Fig.~\ref{figure5} (b)). The demand for top value of efficiency on the transit rate $\Gamma$ is lower than $1 GHz$. Many semiconductor QD materials can satisfy this condition~\cite{Sargent,Sanehira,Kamat}. In addition, due to the dipole moment determines the strength of interaction between electronic system and light, efficiency of energy conversion benefits from large dipole moment as plotted in Fig.~\ref{figure5} (c). The results show efficiency of the inhomogeneous field QD solar cell is not lower than the QD solar cells with p-n type bilayer materials, such as QD-Dye bilayer-sensitized solar cell~\cite{Shalom}, InAs/GaAs quantum dot thin film solar cell~\cite{Tutu,Sogabe}, molecules mediated colloidal QD solar cell~\cite{Baek}, multilayer colloidal QD solar cell~\cite{Lee} and so on.

\begin{center}
\textbf{6. Discussions}
\end{center}

Finally, we discuss about the absorption of sun light. In our model, we just considered single transition energy of QDs, namely $\varepsilon_{2}-\varepsilon_{1}$. At the same time, in the light-matter interaction, light intensity around 1000 Wm$^{-2}$ ($1$ sun) is considered. In fact, sun light consists of electromagnetic field with large range of wavelengths. Therefore, if all QDs in the cell have just one transition as considered in the present model, light effectively absorbed by the cell would be much weaker than the value 1000 Wm$^{-2}$. As a result, current density and efficiency of the photovoltaic system should be very small as manifested in the above figures. Fortunately, there are some direct methods to supplement our present model for higher current and efficiency. For example, practical QDs are usually characterized by multi energy levels, which indicates multi-electron transitions could be allowed actually. In the multiple exciton process, a single photon with large energy can excite two or more electrons in a QD~\cite{Nozik16,Pusch}. In addition, density of QDs considered in our model can be raised by at least $10$ or $10^{2}$ times, which could increase the absorption ability of the QD films.

\begin{center}
\textbf{7. Conclusions}
\end{center}

We proposed a new type photovoltaic effect based on inhomogeneous light intensity in neighboring two QDs system. The inhomogeneous intensity of light in neighboring QDs could lead to polarization of electron number distribution and further induces net current. With the inhomogeneous field photovoltaic effect, materials can be can be symmetrically arranged in solar cells, such as conductor-QDs-conductor, which is more simply structured comparing with the configuration of traditional solar cells which usually rely on p-n type materials for the creation of potential difference. Numerical estimations for a whole solar cell system show that current density and efficiency of this kind of solar cell could be comparable with the traditional solar cells and have large flexibility to be tuned in practice. The inhomogeneous field solar cells mainly require two challenging techniques, one is to create remarkable nonuniform light intensity on the area of a few QDs, the other is to manufacture ultra thin films of QDs. Structure simplicity of such kind photocurrent devices may further decrease the cost of solar cell production in future.

\begin{acknowledgments}
This work was supported by the Scientific Research Project of Beijing Municipal Education Commission (BMEC) under Grant No.KM202011232017.
\end{acknowledgments}

\end{document}